\documentclass[10pt]{amsproc}

\usepackage{amsfonts}
\usepackage{ dsfont }
\usepackage{textcomp}
\usepackage{amssymb}
\usepackage{ upgreek }

\begin{document}


\title[6D Jordan supergravity models]{Six-Dimensional Jordan Supergravity Models}



\author{}

\address{Physics Department, 
University of Athens,15771, Ilisia, Greece}
\curraddr{}
\email{pkouroum@phys.uoa.gr, pkouroum@gmail.com}

\thanks{Based on talk given at StringMath 2011, Philadelphia}

\author []{Peggy Kouroumalou}
\address{}
\curraddr{}
\email{}
\thanks{}

\date{}

\textwidth 12.7cm 
\textheight 16.5cm

\begin{abstract}
In the context of six-dimensional supergravity there is a special class of parent models for five-dimensional theories defined by the four Euclidean simple Jordan algebras of degree 3. We extend this result to include six- dimensional parent models for three infinite families of five-dimensional theories defined by Minkowskian Jordan algebras. Connections of the six-dimensional models  to F-theory constructions are constrained by anomaly cancellation conditions and the   structure of the six-dimensional theory gauge group.

\end{abstract}

\maketitle{}


\bibliographystyle{amsplain}

\section{Introduction}

In the past decades $ \mathcal{N}=2$, $D=5$ supergravity has been studied both independently  \cite {guna,guna2,guna3}
and also as part of the very popular  {\it brane world models} \cite{gherghetta, rattazzi, bagger, peskin, kouroumalou}.That was motivated partly by the fact that certain M-theory compactifications seem to suggest theories which appear five-dimensional 
at a certain intermediate length scale \cite {Ovrut}.

Among the five- dimensional theories, {\it Magical Supergravities} \cite{Gun, Sierra, de Wit} is a very special class whose symmetries correspond to symmetries and underlying algebraic structures of the remarkable geometries of the {\it Magic Square} \cite{Rosenfeld}. In five dimensions these theories describe the coupling of $\mathcal{N}=2$ supergravity to 5,8,14 and 26 vector multiplets and are the unique Maxwell-Einstein supergravity theories (MESGT)  with symmetric target spaces. They are associated to the Euclidean Jordan algebras of degree three  over the four division algebras $\mathds{R}$, $\mathds{C}$, $\mathds{H}$, $\mathds{O}$. In addition, in five dimensions, there exist three infinite families plus an exceptional one  unified MESGT defined by Minkowskian Jordan algebras \cite{gununif}.

  In $D=6$ the magical theories are parent theories from which all magical theories in $D=5$ (and also $D=4,3$) can be obtained by dimensional reduction. 

Since their discovery, higher dimensional and/or  stringy origins of magical supergravity theories have been of great interest. Although higher dimensional constructions  for some of the magical supergravities are known \cite{gun new six, 5,6,7, octonionic,9}, 
a full treatment is obscured by the anomaly cancellation conditions and the (abelian) nature of vectors in the six-dimensional magical theories. We will comment  on  the possibilities of embedding magical supergravities into a higher dimensional framework in more detail, to the last section.

\newpage
\textwidth 12.7cm 
\textheight 21.4cm

Next, we review briefly  the structure of the magical models in $D=5$ and their parent theories in $D=6$. We extend the results 
of \cite{gun new six} to include  parent theories in six dimensions for the Minkowskian Jordan five-dimensional models which are constrained by the anomaly cancellation condition.

In the last  section we review briefly facts and constraints of  higher dimensional proposals for the magical models and comment on new possibilities  based on an F-theory approach \cite{morrison, taylor, kumar}

\section { $\mathcal{N}=2$, $D=5$ Magical Supergravities}

A 5D $\mathcal{N}=2$ MESGT \cite{guna, guna2, guna3} describes the coupling of pure 5D, $\mathcal{N}=2$ supergravity to an arbitrary number, ${\tilde{n}}$, of vector multiplets. The field content of the theory is:

\begin{equation}
\{ e_{\mu}^m ,\Psi_{\mu}^i , A_{\mu}^{\tilde{I}} , \lambda^{i \tilde{a}} , \phi^{\tilde{x}} \}
\end{equation}

\begin{eqnarray}
\tilde{I} \,\,=\,\, 0,1,..., \tilde{n}
\nonumber\\
\tilde{a} \,\,=\,\, 1,..., \tilde{n}
\nonumber\\
\tilde{x} \,\,=\,\, 1,..., \tilde{n} 
\end{eqnarray}

\noindent
where ${ e_{\mu}^m}$ denote the  f$\ddot{u}$nfbein, $\Psi_{\mu}^i$, $i=1,2$ are the two gravitini, $ A_{\mu}^{\tilde{I}}$ are the vector fields of the $\tilde{n}$ vector multiplets and the graviphoton that corresponds to $\tilde{I}=0$, $\lambda^{i \tilde{a}}$ the gaugini and $\phi^{\tilde{x}} $ the scalar fields of the vector multiplet.

The bosonic part of the Lagrangian is:

\begin{eqnarray}
e^{(-1)} \mathcal{L}  = &- &\,\displaystyle \frac{1}{2} R (\omega)  -  \displaystyle\frac{1}{4}\mathring{a}_{\tilde{I}\tilde{J}}F_{\mu\nu}^{\tilde{I}} F^{\tilde{J} \mu\nu} -  \displaystyle\frac{1}{2} g_{\tilde{x}\tilde{y}} (\partial_{\mu}\phi^{\tilde{x}} )  (  \partial^{\mu} \phi^{\tilde{y}} )
\nonumber\\
&+& \displaystyle\frac{e^{(-1)}}{6\sqrt{6}} C_{\tilde{I}\tilde{J}\tilde{K}} \epsilon^{\mu\nu\rho\sigma \lambda} F_{\mu\nu}^{\tilde{I}} F_{\rho\sigma}^{\tilde{J}} A_{\lambda}^{\tilde{K}}
\end{eqnarray}

The $\tilde{n}$ target space $\mathcal{M}$ of the scalar fields  $\phi^{\tilde{x}} $ can be represented as a hypersurfacre
defined by the cubic polynomial 

\begin{equation}
\mathcal {V}(h) =  C_{\tilde{I}\tilde{J}\tilde{K}} h^{\tilde{I}} h^{\tilde{j}} h^{\tilde{K}}  =  1
\end{equation}

\noindent
  in terms of the $\tilde{n} + 1$ real variables  $h^{\tilde{I}} $.
  
  A special class of MESGT's is the Magical Jordan family that corresponds to the four simple Euclidean Jordan algebras of 
  degree 3. These simple Jordan algebras are denoted by $J_3^{\mathds{R}}$, $J_3^{\mathds{C}}$, $J_3^{\mathds{H}}$,$J_3^{\mathds{O}}$

\vspace{-1.5cm}

\begin{eqnarray}
J_3^{\mathds{R}}&:&  \mathcal{M}=SL(3,\mathds{R}) / {SO(3)} \,\,\,\,\, \,\,\,( \tilde{n} = 5 )
\nonumber\\
J_3^{\mathds{C}}&:&  \mathcal{M}=SL(3,\mathds{C}) / {SU(3)} \,\,\,\,\,\,\, \,( \tilde{n} = 8 )
\nonumber\\
J_3^{\mathds{H}}&:&  \mathcal{M}=SU^{*}(6) /  Usp (6) \,\,\,\,\,\,\,\,\,\,( \tilde{n} = 14 )
\nonumber\\
J_3^{\mathds{O}}&:&  \mathcal{M}=E_{6 (-26)} /   F_4 \,\,\,\,\,\,\,\,\,\,\,\,\,\,\,\,\,\,\,\,\,\, ( \tilde{n} = 26 )
\end{eqnarray}

\noindent
where $\tilde{n}$ is the number of vector multiplets. Under the isometry groups $SL(3,\mathds{R}) $ $SL(3,\mathds{C})$,
$ SU^{*}(6)$, $E_{6 (-26)} $ of the scalar manifold $\mathcal{M}$ the 6,9,15,27 vector fields transform irreducibly.

The general element of the Jordan algebra $J_3^{\mathds{A}}$ can be decomposed in terms of the  $J_2^{\mathds{A}}$ subalgebra  as:

\begin{equation}
 \begin{bmatrix}
  x & \Psi  ( {\mathds{A}} ) \\
   \Psi ^{\dagger}  ( {\mathds{A}} ) &  J_2^{\mathds{A}}  \\
 \end{bmatrix}
\end{equation}

\vspace{0.1cm}
\noindent
where $ \Psi  ( {\mathds{A}} )$ is a two-component spinor over  ${\mathds{A}}$, x $\in$ $\mathds{R}$ 

Elements of  $J_3^{\mathds{A}}$  transform as the  $dim( 3+ 3\mathds{A}$)  representation of $SL( (3,\mathds{A} ) $ while  $J_2^{\mathds{A}}$ transform as the  $dim( 2+\mathds{A}$ ) of  $SL( (2,\mathds{A}) $ that is the  {\bf{10}, \bf{6}, \bf{4}, \bf{3}}
of $SO( 9,1 )$, $SO( 5,1 )$, $SO( 3,1 )$ and $SO( 2,1 )$ respectively.

\section{Minkowskian Jordan algebras}

Non compact analogs  $J_{ (q, n-q)}^{\mathds{A}}$ of the Euclidean Jordan algebras $J_n^{\mathds{A}}$ over the division algebras   ${\mathds{R}}$, $\mathds{C}$, $\mathds{H}$  for $n \geqslant 3$ and of   $J_3^{\mathds{O}}$  are realized by  matrices $X$  that are hermitean with respect to a non- Euclidean metric $(q, n-q)$:

\begin{equation}
( \eta X )^{\dagger} = \eta X , \,\,\, \forall  X \in J_{ (q, n-q)}
\end{equation}
                                                   
\noindent
We consider Minkowskian Jordan algebras  $J_{(1,N)}^{\mathds{A}}$ of degree $n = N+1$ with $\eta$ the Minkowski metric
$\eta = ( -,+,+,....,+ )$. The general element of  $J_{(1,N)}^{\mathds{A}}$ is of the form

\[
\begin{bmatrix}
  x &  -Y^{\dagger}\\
  Y   &  J_N^{\mathds{A}}  \\
 \end{bmatrix}
\]

\noindent
where $x \in \mathds{R}$, $Y $ is an N-column vector of $\mathds{A}$  and  $J_N^{\mathds{A}}$ is the N-dimensional Euclidean Jordan algebra. Under the automorphism group $Aut J_{(1,N)}^{\mathds{A}}$ the $ J_{(1,N)}^{\mathds{A}}$ decompose as:

\begin{equation}
J_{(1,N)}^{\mathds{A}} = {\bf{1}} \oplus  \{{traceless \,elements}\}
\end{equation}

\noindent
Denoting the traceless elements of  $J_{(1,N)}^{\mathds{A}}$ as $T_{\tilde{I}}$, $\tilde{I} = 0,1,..., (D-2)$  with $D = dim  J_{(1,N)}^{\mathds{A}}$, the d-symbols are given by

\begin{equation}
d_{\tilde{I}\tilde{J}\tilde{K}} \equiv d_{\tilde{J}\tilde{K}}^{\tilde{L}} \tau_{\tilde{L}\tilde{I}} ,\,\,\,\,\,\,  \tau_{\tilde{L}\tilde{I}} = tr ( T_{\tilde{L}} \circ T_{\tilde{I}} )
\end{equation}

\noindent
The normal Jordan product is denoted by $\circ$:  $\, A\, \circ\, B = \frac{1}{2} ( AB + BA)$. We must note here that the traceless elements of the Minkowskian Jordan algebras  $J_{(1,N)}^{\mathds{A}}$ do not close under  $\circ$ but under  a symmetric product  $\star$, defined as:

\begin{equation}
A \star B : =  A \circ B - \displaystyle\frac{1}{ ( N+1 )} tr ( A  \circ  B ) \,  {\bf{1}}
\end{equation}

\noindent
By identifying the d-symbols of the traceless elements of the  $J_{(1,N)}^{\mathds{A}}$ with the $C_{\tilde{I}\tilde{J}\tilde{K}}$  (2.4) of a MESGT :   $d_{\tilde{I}\tilde{J}\tilde{K}}  =  C_{\tilde{I}\tilde{J}\tilde{K}}$ we obtain a unified MESGT ie: all vector fields transform irreducibly under the $Aut  ( J_{(1,N)}^{\mathds{A}} )$ \cite{gununif}.

There exist three infinite families of Minkowskian Jordan algebras for  $\mathds{R} $, $\mathds{C}$, $ \mathds{H}$ ($ N \geq 2$) and one single MESGT based on the octonionic Minkowskian Jordan algebra $J_{(1,2)}$. 
In Table 1,we list the Minkowskian Jordan  algebras  as they  appear in \cite{gununif}. D denotes the dimension of the Jordan Minkowskian algebra,
$\tilde{ n}$ is  the number of vector fields $\tilde{ n}$ : $ D = \tilde{n} + 2 $.

The choice of the Minkowskian signature of the metric $\eta = ( +,-,-,...- )$ ensures the positivity of the quantities $\mathring{a}_{\tilde{I}\tilde{J}}$ and   $ g_{\tilde{x}\tilde{y}}$ which in their turn ensure the positivity of the kinetic terms of the vector fields and scalars, leading to a physically acceptable MESGT.

\begin{center}
 \begin{table}
  \caption {Minkowskian Jordan algebras }
  \begin{tabular}{| l  | c | r |  r | }
    \hline \hline
    $\,\,\,$  J  &  D &  Aut ( J ) & $\tilde{n}$    vector fields \\ \hline
     $ J_{( 1, N )}^{\mathds{R}}$  &  $  \displaystyle\frac{1 } {2} (N + 1) ( N + 2) $  &  $ SO (N, 1)$   &  $  \displaystyle\frac{1 } {2}  N (N + 3) \,\,\,\,\,\, $  \\ 
   $ J_{(1, N )}^{\mathds{C}}$  &  ${ (N + 1) }^2$    &   $ SU (N, 1) $ & $ N (N + 2)\,\,\,\,\,\,\,$\\  
  $ J_{(1, N )}^{\mathds{H}}$  &   $ ( N + 1) ( 2N + 1) $  &  $USp(2N,2)$ & $ N(2N + 3)\,\,\,\,\,\,\,$\\
  $ J_{(1, 2)}^{\mathds{O}}$   &  27  &   $ F_{4(-20)}\,\,\,\,\,$ &  26  \,\,\,\,\,\,\,\,\,\,\,\,\,\,\,\,\,\,\,\\ \hline
 \end{tabular}
\end{table}
\end{center}

\vspace{-0.6cm}
Three of the  above Minkowskian Jordan algebras are equivalent  to the three magical MESGT's:  $J_{(1,2)}^{\mathds{R}}$, $J_{(1,2)}^{\mathds{C}}$, $J_{(1,2)}^{\mathds{H }} $ are equivalent to $J_3^{\mathds{C}}$, $J_3^{\mathds{H}}$, $J_3^{\mathds{O}}$ respectively. In that case, the cubic polynomials $\mathcal{V}(h) $  of the equivalent theories agree \cite{gununif,allison}.

\section {$\mathcal{N} = (1,0)$ Six Dimensional Supergravity}
 
 \subsection{ $\mathcal{N} = (1,0)$ six dimensional supergravity field content}

 The  field content of the chiral $\mathcal{N}=(1,0)$ supergravity coupled to $n_T$ tensor multiplets, $n_V$ vector multiplets and $n_H$ hypermultiplets \cite{nishino,sezgin,romans,riccioni} is the following:

\vspace{0.4cm} 

\begin{center} 
\begin{tabular} {l  c }
$\{e_{\mu}^m, B_{\mu\nu}^+, \Psi_{\mu L}^A  \}$  $\,\,\,$  & supergravity  multiplet\\
\\
$\{ B_{\mu\nu}^ { - M}, \phi^M, \chi_R^{A M}     \}$ &  $\,\,\,$ tensor multiplet \\
\\
$\{ A_{\mu}^x,  {\lambda }_R^{A   x}   \}$ & $\,\,\,$  vector multiplet\\
\\
$\{ \Psi_R^{\upalpha}, {\phi}^{\alpha}      \}$   & $\,\,\,$  hypermultiplet\\
\end{tabular}
\end{center}

\vspace{0.4cm}
\noindent
with  $M = 1,...., n_T$, $ x = 1,...., n_V$,  $\upalpha = 1,...., 2n_H$,  $\alpha = 1,..., 4n_H$ and $ A = 1,2$ is the $USp(2)$ index of the fermions which are symplectic Majorana. The $n_T$ scalars in the tensor multiplets  parametrize the coset space $SO (1, n_T)/SO(n_T)$. The vectors and the gauginos are in the adjoint representation of the gauge group. The scalars $\phi^M$  
parametrize a quaternionic manifold $\mathcal{M_Q}$.

\subsection{6D parent theories of  the magical supergravities}

The magical supergravities defined in section 2, can be truncated to theories belonging to the generic Jordan family generated  by reducible Jordan algebras  $ ( \mathds{R} \oplus  J_2^{\mathds{A}}  )$ \cite{pavlyk,gun new six}.
The isometry groups of the scalar manifolds of the truncated  5D theories are:

\vspace{0.7cm}
\begin{center}
\begin{equation}
\begin{tabular}{l c}
    $\mathds{R}$   $\oplus$  $J_2^{\mathds{R}}$  $ \rightarrow $ &  SO(1,1) $\times$ $SL(2, \mathds{R} )$ $\subset$  $SL(3, \mathds{R} )$\\
 \\
 $\mathds{R}$   $\oplus$  $J_2^{\mathds{C}}$  $ \rightarrow $ &  SO(1,1) $\times$ $SL(2, \mathds{C} )$ $\subset$  $SL(3, \mathds{C} )$\\
 \\   
$\mathds{R}$   $\oplus$  $J_2^{\mathds{H}}$  $ \rightarrow $ &  SO(1,1) $\times$ $SL(2, \mathds{H} )$ $\subset$  $ SU^{\star}(6) $\\
 \\
$\mathds{R}$   $\oplus$  $J_2^{\mathds{O}}$  $ \rightarrow $ &  SO(1,1) $\times$ $SL(2, \mathds{O} )$ $\subset$  $ E_ {6(-26)} $\\
 \\
\end{tabular}
\end{equation}
\end{center}

\noindent
In the decomposition (2.6)  of $J_3^{\mathds{A}}$,    the tensor fields correspond to the elements of $J_2^{\mathds{A}}$ and the vector fields to elements of the form

\begin{equation}
 \begin{bmatrix}
  0 & \Psi  ( {\mathds{A}} ) \\
   \Psi ^{\dagger}  ( {\mathds{A}} ) &  0  \\
 \end{bmatrix}
\end{equation}

\noindent
with $\Psi( \mathds{A} )$ a two-component spinor over $\mathds{A}$. 
The truncated theories (4.1) descend from six dimensions with $n_T  = 2,3,5$ and 9 tensor multiplets and no vector multiplets
The scalar target spaces of the four magical supergravities in six dimensions are $SO( 2,1 )/SO( 2 )$, $SO( 3,1 )/SO( 3 )$, $SO( 5,1 )/SO( 5 )$ and $SO( 9,1 )/SO( 9 )$. The vectors in these theories transform in the spinor representation of
$SO( 1, n_T )$.

In order to cancel gravitational anomalies  $n_H$ hypermultiplets must be included so that the well known relation

\begin{equation}
n_H =  273 + n_V -29 n_T
\end{equation}

\noindent 
is fulfilled. For the magical supergravities this condition gives the values $ (n_T, n_V, n_H )$ for the four models: $ ( 2, 2, 217 )$, $(  3, 4, 190 )$, $ (  5, 8, 136 )$ and $( 9, 16, 28 )$.

 The above condition ensures the vanishing of the term $ \sim tr R^4$. We will discuss in more detail the anomaly cancellation in the last section.

\subsection {6D parent theories of the Minkowskian Jordan supergravities}

The procedure we follow to obtain the parent theories in six dimensions of the five-dimensional  Minkowskian MESGT's, is identical to the one used in the previous section to obtain the parent theories of the magical supergravities.

\vspace{0.2cm}
 $ \bf  J_{ (1, N)}^{\mathds{R}}$

The traceless element of  $J_{ (1, N)}^{\mathds{R}}$  is of the form:
 \[\begin{bmatrix}
  x &  -Y^{\dagger}\\
  Y   &  J_N^{\mathds{R}}  \\ 
 \end{bmatrix}
\]  

\vspace{0.2cm} 
$\bullet$ dim(traceless element of $J_{ (1, N)}^{\mathds{R}}$) = $\displaystyle\frac{1}{2} N ( N + 3 )$

$\bullet$ number of vector fields : N

$\bullet$ number of tensor multiplets : $\displaystyle\frac{ N(N + 1)}{2}$ - 2

 $ \bf  J_{ (1, N)}^{\mathds{C}}$

The traceless element of  $J_{ (1, N)}^{\mathds{C}}$  is of the form:
 \[\begin{bmatrix}
  x &  -Y^{\dagger}\\
  Y   &  J_N^{\mathds{C}}  \\
 \end{bmatrix}
\]

$\bullet$ dim(traceless element of $J_{ (1, N)}^{\mathds{C}}$) = $ N ( N + 2 )$

$\bullet$ number of vector fields :  N

$\bullet$ number of tensor multiplets : $ N^2 - 2$

\vspace{0.5cm}
$ \bf  J_{ (1, N)}^{\mathds{H}}$

The traceless element of  $J_{ (1, N)}^{\mathds{H}}$  is of the form:
 \[\begin{bmatrix}
  x &  -Y^{\dagger}\\
  Y   &  J_N^{\mathds{H}}  \\
 \end{bmatrix}
\]  

\vspace{0.1cm}
$\bullet$ dim(traceless element of $J_{ (1, N)}^{\mathds{H}}$) = $ N ( 2N + 3 )$

$\bullet$ number of vector fields : 4N

$\bullet$ number of tensor multiplets : $ 2N^2 - N - 2$

$ \bf  J_{ (1, 2)}^{\mathds{O}}$

$\bullet$ number of vector fields : 8

$\bullet$ number of tensor multiplets : 4

\noindent
Anomaly cancellation condition, (4.3), specifies the number of hypermultiplets in each of the series of the above models:

\vspace{0.2cm}
 $ \bf  J_{ (1, N)}^{\mathds{R}}$

$\bullet$ $n_H = 331 + N -\displaystyle\frac{29}{2} N(N + 1)$  $\longrightarrow$  $N \leq 4$

\vspace{0.2cm}
 $ \bf  J_{ (1, N)}^{\mathds{C}}$
 
$\bullet$ $n_H = 331 + 2N - 29 N^2$  $\longrightarrow$  $N \leq 3$ 

\vspace{0.2cm}
 $ \bf  J_{ (1, N)}^{\mathds{H}}$
 
$\bullet$ $n_H = 331 + 33N - 58 N^2$  $\longrightarrow$  $N =2 $ 

\vspace{0.2cm}
 $ \bf  J_{ (1, 2)}^{\mathds{O}}$
 
$\bullet$ $n_H = 165$  
      
\noindent
Among the above models the one corresponding to  $J_{ (1, 2)}^{\mathds{O}}$ accomodates the maximum number of vectors:
 $ \bf  J_{ (1, 2)}^{\mathds{O}}$ $\longrightarrow$ $n_V = 8$,  $n_T = 4$, $n_H = 165$, while there is a model with one tensor multiplet:  $ \bf  J_{ (1, 2)}^{\mathds{R}}$ $\longrightarrow$ $n_V = 2$,  $n_T = 1$, $n_H = 246$.

The tensor and vector fields in the 6D models transform in the vector and spinor representation of $SO( 1, n_T )$, respectively. As we will see in the next section this is very restrictive for the allowed gauge group.

\section{Discussion - Open Problems}

We reviewed the  magical supergravities in five and six dimensions, defined by the four simple Euclidean Jordan algebras of degree 3. These correspond in five dimensions to $\mathcal{N} = 2$ MESGT's in which all the vector fields  transform irreducibly under a global symmetry group of the Lagrangian and   whose target manifolds are symmetric spaces. 

In addition,  we reviewed the MESGT's defined by the  Minkowskian Jordan algebras. We extended the results of \cite{gun new six} to include parent 6D models of the Minkowskian  Jordan MESGT;s. The anomaly cancellation condition (4.3)
restricts the field content  and limits the possible models to $J_{(1,4)}^{\mathds{R}}$, $J_{(1,3)}^{\mathds{R}}$, $J_{(1,2)}^{\mathds{R}}$, $J_{(1,3)}^{\mathds{C}}$, $J_{(1,2)}^{\mathds{C}}$, $J_{(1,2)}^{\mathds{H}}$ and     $J_{(1,2)}^{\mathds{O.}}$. A full analysis of  these models will appear elsewhere  \cite{peggnew} together with an analysis of the magical models in 6D (including gauged models \cite{gun new six})  from the point of view  of anomaly cancellation. More
specifically, in six dimensions, irreducible anomalies $\sim tr R^4$, $\sim F^4$ have to be absent. 
\noindent
Once these conditions are  satisfied, the  anomaly polynomial  contains the following terms \cite{erler,honecker}:

\[
I_8  \sim ( n_H - n_V - 7n_T + 51) ( tr R^2 )^2 + B ( tr R^2 )( tr F^2 ) + C ( tr F^2 )^2 + D ( tr F^3 ) ( trF )
\]

\noindent 
For the case of the magical octonionic model, the first term yields $n_T - 9$ which vanishes identically for $n_T = 9$ while for the rest of the models this term is present. The rest of the anomalies can be treated with a suitable type Green-Schwarz mechanism \cite{peggnew}.

Stringy origins of magical supergravities have been of great interest. Although stringy constructions of some of them are known with or without hypermultiplets  \cite{gun new six, 5,6,7, octonionic,9},  a full treatment is missing. The main obstacle to this is the abelian nature of the vector fields which transform in the spinor representation of $SO(1, n_T)$. On the other hand, F-theory compactifications on Calabi-Yau threefolds provide a large class of $\mathcal{N}=(1, 0)$ 6D models \cite{Vaf1, Vaf2}.
Recently, a systematic analysis of  anomaly free 6D models in the context of F-theory has been developed \cite{kumar, taylor, morrison}. In this analysis every consistent 6D theory is associated with an integral lattice that is determined by the anomaly
constraints. Next, data of the integral lattice are mapped to topological data of F-theory. Not all of the low-energy consistent 6D models  seem to have an F-theory realization. Since the gauge groups of the 6D models considered in \cite{kumar, taylor, morrison} were  assumed to contain no abelian factors, the realization of the Jordan 6D models in this context seems to be impossible. 

At the time of writing this work,  two interesting  works \cite{park}, \cite{taylorpar}  towards the direction of embedding     $\mathcal{N}=(1, 0)$  6D supergravity models with abelian  factors in F-theory,  drew our attention. In these  papers, anomaly constraints are analyzed and the number of abelian factors is bounded in terms of tensor multiplets for purely abelian theories. It will be interesting to check the special class of Jordan supergravities under this framework.


\vspace{2cm}

  
  
\begin{thebibliography}{99}

\bibitem{guna}
M.~Gunaydin, M.~Zagermann,
  Nucl.\ Phys.\  {\bf B572}, 131-150 (2000).
  [hep-th/9912027].
\bibitem{guna2}
M. Gunaydin,  G. Sierra, P. K. Townsend, Phys. Lett. {B133},72 (1983);  Class. Quantum Grav. {3}  763 (1986);  Phys. Rev. Lett.
{B53}, 332 (1984); Phys. Lett. {B144}, 41 (1984).

\bibitem{guna3}
M.~Gunaydin, M.~Zagermann,
  Phys.\ Rev.\  {\bf D62}, 044028 (2000)
  [hep-th/0002228].


\bibitem{gherghetta}

T. Gherghetta, Antonio Riotto, Nucl. Phys. {{B623}} ,  97 (2002).

\bibitem{rattazzi} 

R.~Rattazzi, C.~A.~Scrucca and A.~Strumia,
  Nucl.\ Phys.\  B {74} (2003) 171
 
  [arXiv:hep-th/0305184].
 




\bibitem{peskin}

E. A. Mirabelli  and  Peskin, Phys. Rev. {{D58}} 0605002  (1998).
\bibitem{bagger}

J. A.Bagger, F.Feruglio  and F. Zwirner, Phys. Rev. Lett. {{88}}, 101601 (2002)

[arXiv: hep-th/0107128 ];  J. A. Bagger,  F.Feruglio  and  F.Zwirner,   JHEP  { {0202}}, 010 (2002) [arXiv: hep-th/0108010 ].

\bibitem{kouroumalou}
G.A.Diamandis, B.C.Georgalas, P.Kouroumalou, A.B.Lahanas,

[arXiv:hep-th/1001.5382] Phys.Lett.B692:26-3, 2010.

G.A.Diamandis, B.C.Georgalas, P.Kouroumalou, A.B.Lahanas,

[arXiv:hep-th/0711414]  Phys. Let.B 660:247-253, 2008.

G.A.Diamandis, B.C.Georgalas, P.Kouroumalou, A.B.Lahanas,

[arXiv:hep-th/0402228] Phys. Let.B 602:112-122, 2004.
\bibitem{Ovrut}
A.~Lukas, B.~A.~Ovrut, K.~S.~Stelle and D.~Waldram,
  Nucl.\ Phys.\  B {\bf 552}, 246 (1999)
  
  [arXiv:hep-th/9806051].






\bibitem{Gun}
M. Gunaydin, G. Sierra,   P. K. Townsend Nucl. Phys. B242 (1984) 244. 
\bibitem{Sierra}
 M. Gunaydin, G. Sierra,   P. K. Townsend, Nucl. Phys. B253 (1985) 573. 
\bibitem{de Wit}
 M.Gunaydin, G. Sierra, and P. K. Townsend, Phys.Lett.B133(1983) 72.




\bibitem{Rosenfeld}

H. Freudenthal, Nederl. Akad. Wetensch. Proc. A62 (1959), 447; 
B.A. Rozenfeld, Dokl. Akad. Nauk. SSSR 106 (1956),600; 
J. Tits, Mem. Acad. Roy. Belg. Sci. 29 (1955) fasc.3. 

\bibitem{gununif}

M.Gunaydin,  M.Zagermann,
  JHEP {\bf 0307}, 023 (2003)
  [arXiv:hep-th/0304109].

\bibitem{gun new six}

M.Gunaydin, H.Samtleben,  E.Sezgin,
  Nucl.\ Phys.\  B {\bf 848}, 62 (2011)
  [arXiv:1012.1818 [hep-th]].

\bibitem{5}

A. Sen, C. Vafa,  Nucl. 
Phys. B455 (1995) 165–187, hep-th/9508064 - 
 9. 
  
\bibitem{6}

M. Gunaydin. talk titled ” From d=6, N=1 to d=4, N=2, No-scale models and 
Jordan Algebras ” at the Conference 30 Years of Supergravity in Paris, 
October, 2006 
\bibitem{7}

Y. Dolivet, B. Julia, and C. Kounnas,JHEP 02 (2008) 097, 0712.2867.4 
49. 

\bibitem{octonionic}

M. Bianchi and S. Ferrara,
models,” JHEP 02 (2008) 054, 0712.2976.

\bibitem{9}
A. N. Todorov, “CY manifolds with locally symmetric moduli spaces,” 
arXiv:0806.4010.


\bibitem{morrison}

V.~Kumar, D.~R.~Morrison and W.~Taylor,
  JHEP {\bf 1011}, 118 (2010)
  [arXiv:1008.1062 [hep-th]].

\bibitem{taylor}

V.~Kumar, D.~R.~Morrison and W.~Taylor,
  JHEP {\bf 1002}, 099 (2010)
  [arXiv:0911.3393 [hep-th]].

\bibitem{kumar}

V. Kumar and W. Taylor,  JHEP {\bf 0912}, 050 (2009) 
[arXiv:0910.1586 [hep-th]]. 


\bibitem{nishino}

H. Nishino and E. Sezgin,  Nucl. Phys. B278 (1986) 353–379. 
\bibitem{sezgin}

H. Nishino and E. Sezgin,
Nucl. Phys. B505 (1997) 497–516, hep-th/9703075. 

\bibitem{romans}
S. Ferrara, F. Riccioni, and A. Sagnotti,Nucl. Phys. B519 (1998) 115–140, 
hep-th/9711059. 

\bibitem{riccioni}

F. Riccioni, Nucl. Phys. B605 (2001) 245–265, hep-th/0101074.


\bibitem{pavlyk}

 M.~Gunaydin, O.~Pavlyk,
  JHEP {\bf 0508}, 101 (2005).
  [hep-th/0506010].


\bibitem{allison}
B.N. Allison and J.R. Faulkner  Trans. Amer. Math. Soc. 283 (1984) 185-210. 


\bibitem{peggnew}

P. Kouroumalou: to appear.


\bibitem{erler}

J. Erler, J. Math. Phys. 35, 1819 (1994) 
arXiv:hep-th/9304104. 

\bibitem{honecker}
G.~Honecker,
  Nucl.\ Phys.\  {\bf B748}, 126-148 (2006).
  [hep-th/0602101].
\bibitem{Vaf1}
D.~R.~Morrison, C.~Vafa,
  Nucl.\ Phys.\  {\bf B473}, 74-92 (1996).
  [hep-th/9602114].
\bibitem{Vaf2}
D.~R.~Morrison, C.~Vafa,
  Nucl.\ Phys.\  {\bf B476}, 437-469 (1996).
  [hep-th/9603161].
\bibitem{taylorpar}
D. S. Park , W. Taylor
"Constraints on 6D Supergravity Theories 
with Abelian Gauge Symmetry "
[hep-th  11105916]

\bibitem{park}
D. S. Park  ``Anomaly Equations and Intersection Theory ''
[hep-th 1111.2351]
\end{thebibliography}
\end{document}